\documentclass[sigconf,natbib=true]{acmart}
\usepackage{multirow}
\usepackage{subcaption}
\usepackage{color}
\usepackage{enumitem}

\AtBeginDocument{%
  \providecommand\BibTeX{{%
    \normalfont B\kern-0.5em{\scshape i\kern-0.25em b}\kern-0.8em\TeX}}}

\setcopyright{acmlicensed}
\copyrightyear{2018}
\acmYear{2018}
\acmDOI{XXXXXXX.XXXXXXX}

\acmConference[SIGIR '25]{Make sure to enter the correct
  conference title from your rights confirmation emai}{July 13--17,
  2025}{Padua, Italy}
\acmBooktitle{Woodstock '18: ACM Symposium on Neural Gaze Detection,
 June 03--05, 2018, Woodstock, NY} 
\acmISBN{978-1-4503-XXXX-X/18/06}

\setlist[itemize]{leftmargin=*}

\begin{document}

%%
%% The "title" command has an optional parameter,
%% allowing the author to define a "short title" to be used in page headers.

\title{Data Augmentation as Free Lunch: Exploring the Test-Time Augmentation for Sequential Recommendation}

%%
%% The "author" command and its associated commands are used to define
%% the authors and their affiliations.
%% Of note is the shared affiliation of the first two authors, and the
%% "authornote" and "authornotemark" commands
%% used to denote shared contribution to the research.
\author{Yizhou Dang}
\affiliation{%
  \institution{Software College, Northeastern University}
  \city{Shenyang}
  \country{China}
}
\email{dangyz@stumail.neu.edu.cn}

\author{Yuting Liu}
\affiliation{%
  \institution{Software College, Northeastern University}
  \city{Shenyang}
  \country{China}
}
\email{yutingliu@stumail.neu.edu.cn}

\author{Enneng Yang}
\affiliation{%
  \institution{Software College, Northeastern University}
  \city{Shenyang}
  \country{China}
}
\email{ennengyang@stumail.neu.edu.cn}

\author{Minhan Huang}
\affiliation{%
  \institution{Software College, Northeastern University}
  \city{Shenyang}
  \country{China}
}
\email{huangminhan@stumail.neu.edu.cn}

\author{Guibing Guo}
\authornotemark[1]
\affiliation{%
  \institution{Software College, Northeastern University}
  \city{Shenyang}
  \country{China}
}
\email{guogb@swc.neu.edu.cn}

\author{Jianzhe Zhao}
\authornote{Corresponding authors.}
\affiliation{%
  \institution{Software College, Northeastern University}
  \city{Shenyang}
  \country{China}
}
\email{zhaojz@swc.neu.edu.cn}

\author{Xingwei Wang}
\affiliation{%
  \institution{School of Computer Science and Engineering, Northeastern University}
  \city{Shenyang}
  \country{China}
}
\email{wangxw@mail.neu.edu.cn}

%%
%% By default, the full list of authors will be used in the page
%% headers. Often, this list is too long, and will overlap
%% other information printed in the page headers. This command allows
%% the author to define a more concise list
%% of authors' names for this purpose.
\renewcommand{\shortauthors}{Yizhou Dang et al.}

%%
%% The abstract is a short summary of the work to be presented in the
%% article.
\begin{abstract}
Data augmentation has become a promising method of mitigating data sparsity in sequential recommendation. Existing methods generate new yet effective data during model training to improve performance. However, deploying them requires retraining, architecture modification, or introducing additional learnable parameters. These steps are time-consuming and costly for well-trained models, especially when the model scale becomes large. In this work, we explore the test-time augmentation (TTA) for sequential recommendation, which augments the inputs during the model inference and then aggregates the model's predictions for augmented data to improve final accuracy. It avoids significant time and cost overhead from the previously mentioned steps. We first experimentally disclose the potential of existing augmentation operators for TTA and find that the Mask and Substitute consistently achieve better performance. Further analysis reveals that these two operators are effective because they retain the original sequential pattern while adding appropriate perturbations. Meanwhile, we argue that these two operators still face time-consuming item selection or interference information from mask tokens. Based on the analysis and limitations, we present TNoise and TMask. The former injects uniform noise into the original representation, avoiding the computational overhead of item selection. The latter blocks mask token from participating in model calculations or directly removes interactions that should have been replaced with mask tokens. Comprehensive experiments demonstrate the effectiveness, efficiency, and generalizability of our method. Our codes is available at \url{https://github.com/KingGugu/TTA4SR}.

\end{abstract}

%%
%% The code below is generated by the tool at http://dl.acm.org/ccs.cfm.
%% Please copy and paste the code instead of the example below.
%%


%%
%% Keywords. The author(s) should pick words that accurately describe
%% the work being presented. Separate the keywords with commas.

%% A "teaser" image appears between the author and affiliation
%% information and the body of the document, and typically spans the
%% page.
% \begin{teaserfigure}
%   \includegraphics[width=\textwidth]{sampleteaser}
%   \caption{Seattle Mariners at Spring Training, 2010.}
%   \Description{Enjoying the baseball game from the third-base
%   seats. Ichiro Suzuki preparing to bat.}
%   \label{fig:teaser}
% \end{teaserfigure}

% \received{20 February 2007}
% \received[revised]{12 March 2009}
% \received[accepted]{5 June 2009}

%%
%% This command processes the author and affiliation and title
%% information and builds the first part of the formatted document.
\maketitle

\section{Introduction}
Sequential recommendation (SR) has received much attention due to its well-consistency with real-world recommendation situations \cite{zhai2025combinatorial}. Over the past few decades, many SR models have made significant achievements in learning the transition patterns and user preferences from historical sequences \cite{tang2018personalized,hidasi2015session,TTTRec,meng2023parallel}. Since most users tend to interact with only a few items on the platform, the widespread problem of data sparsity limits the performance of these models \cite{dang2024data}. For this reason, researchers have proposed many data augmentation methods to mitigate this phenomenon \cite{liu2021contrastive,dang2024repeated}.

Earlier work used heuristic methods to increase the training data, such as Sliding Windows \cite{tang2018personalized}. Later, some researchers found that heuristics tend to produce data of poorer quality and sometimes even impair model performance \cite{dang2023ticoserec}. To tackle this, they proposed model-based augmentation methods to generate high-quality augmented data by counterfactual thinking \cite{wang2021counterfactual}, diffusion models \cite{liu2023diffusion} or bi-directional transformer \cite{liu2021augmenting}. These methods usually require training specialized data augmentation modules to generate new data based on existing data \cite{dang2024repeated}. With the success of contrastive learning, many sequence data augmentation operators have been proposed to construct views for contrastive learning \cite{liu2021contrastive,xie2022contrastive}.

\begin{figure}[!t]
  \centering
  \includegraphics[scale=0.50]{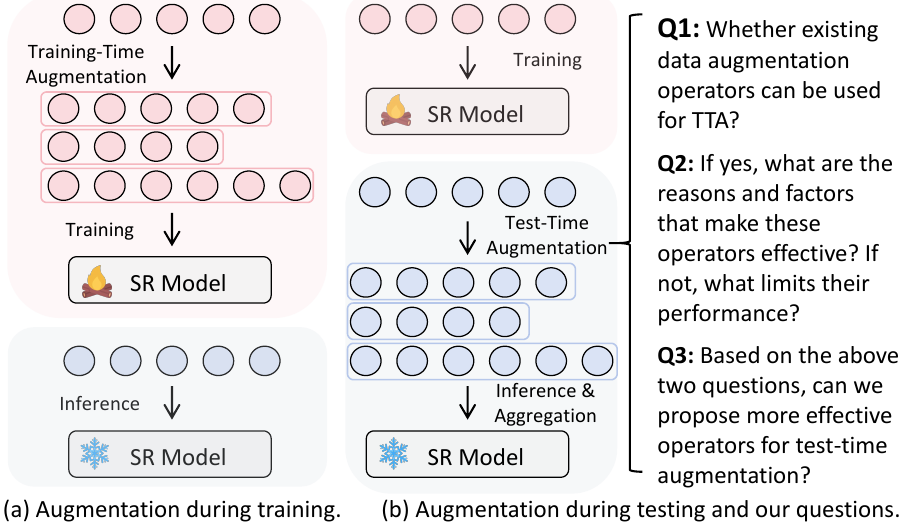}
  \vspace{-0.5em}
  \caption{Illustration of training-time augmentation and test-time augmentation. We also present our research questions.}
  \label{fig:example}
  \vspace{-1em}
\end{figure}

However, existing augmentation methods typically augment sequences during model training. Although effective, deploying them requires retraining the learned model. The newly generated data can significantly extend the time required for retraining or training new models from scratch. In addition, some model-based augmentation methods require modifying the training process or model structure \cite{dang2024repeated}. In real-world scenarios, these steps come with significant additional costs \cite{zhang2022memo}. The problem is further exacerbated for large-scale datasets or models. In facing these challenges, test-time augmentation provides a feasible option \cite{kimura2021understanding,jin2022empowering}. As illustrated in Figure \ref{fig:example}, it augments the inputs during the model inference and then aggregates the model's predictions for augmented data to improve final accuracy. Since there is no need to retrain or change the original model structure, TTA avoids high cost and time overheads. 

In this work, inspired by the success of TTA in the field of computer vision \cite{shanmugam2021better,kim2020learning}, we explore the test-time augmentation for sequential recommendation and try to answer the following three gradually deepening research questions:

\begin{itemize}
\item \textbf{Q1}: Whether existing sequence data augmentation operators can be used for test-time augmentation?
\item \textbf{Q2}: If yes, what are the reasons and factors that make these operators effective? If not, what limits their performance?
\item \textbf{Q3}: Based on the above two questions, can we propose more effective operators for test-time augmentation?
\end{itemize}

To answer the above questions, we first conduct an empirical study to test the performance of representative existing operators when used for TTA. These operators improve the performance of the original model. Among them, the Mask and Substitute consistently perform better (\textbf{Q1}). Further, we analyzed these operators in terms of data similarity between augmented and original data and the effect that these operators have on the original sequential patterns. We find that Mask and Substitute perform better because they introduce appropriate perturbations while preserving the original sequential patterns. In contrast, the other operators may destroy the original sequential patterns or lose important recent interactions. In addition, we leverage large language models \cite{zhang2024finerec,lin2023can} to identify the key interactions of the original sequences. Based on the identification results, we explore the impact of the position selecting mode of operators on the performance. It turns out that the random selection method is a satisfactory scheme (\textbf{Q2}).

However, we argue that the Substitute operator is limited by the high computational and time overhead in selecting similar items. Also, the mask tokens used by the Mask introduce interfering information during the inference phase, impairing the final performance. Based on the previous analysis and the limitations of these two operators, we present our test-time augmentation operators, TNoise and TMask. The TNoise injects uniform noise into the original representation, avoiding the time and computational overhead of item selection. Meanwhile, injection noise can introduce appropriate perturbations while preserving the original sequential patterns. The TMask blocks mask tokens from participating in model calculations or directly removes interactions that should have been replaced with mask tokens, making the high-order item relationship in sequence available for the model and avoiding interfering information from mask tokens (\textbf{Q3}). We conduct extensive experiments on widely used datasets, representative sequential models, and augmentation methods. The results demonstrate the superiority of our method in terms of effectiveness, efficiency, and generalizability.

In summary, our work makes the following contributions:
\begin{itemize}
  \item We explore the potential of existing sequence data augmentation operators when used for test-time augmentation. The results show that Mask and Substitutes consistently perform better.

  \item We analyze why Mask and Substitute outperform other operators from multiple perspectives. They retain the original sequential pattern while adding appropriate perturbations.

  \item Based on the analysis and limitations, we propose two new test-time augmentation operators, TNoise and TMask, which inject uniform noise and avoid interfering information.
  
  \item We conduct comprehensive experiments on multiple datasets, sequential models, and augmentation methods, demonstrating the effectiveness, efficiency, and generalizability of our method.
\end{itemize}

\section{Preliminaries}
\subsection{Problem Formulation}

Suppose we have user set $\mathcal{U}$ and item set $\mathcal{V}$. Each user $u \in \mathcal{U}$ is associated with a sequence of interacted items in chronological order $s_u=[v_1, \ldots, v_j, \ldots, v_{\left|s_u\right|}]$, where $v_j \in \mathcal{V}$ indicate the item that user $u$ has interacted with at time step $j$ and $\left|s_u\right|$ is the sequence length. Given the sequences of interacted items $s_u$, sequential recommendation aims to accurately predict the most possible item $v^{*}$ that user $u$ will interact with at time step $\left|s_u\right|+1$, formulated as:
\begin{equation}
    \underset{v^{*} \in \mathcal{V}}{\arg \max} \;\; P\left(v_{\left|s_u\right|+1}=v^{*} \mid s_u \right).
\end{equation}
This equation can be interpreted as calculating the probability of all candidate items and selecting the highest one for recommendation.

\begin{figure}[!t]
  \centering
  \includegraphics[scale=0.35]{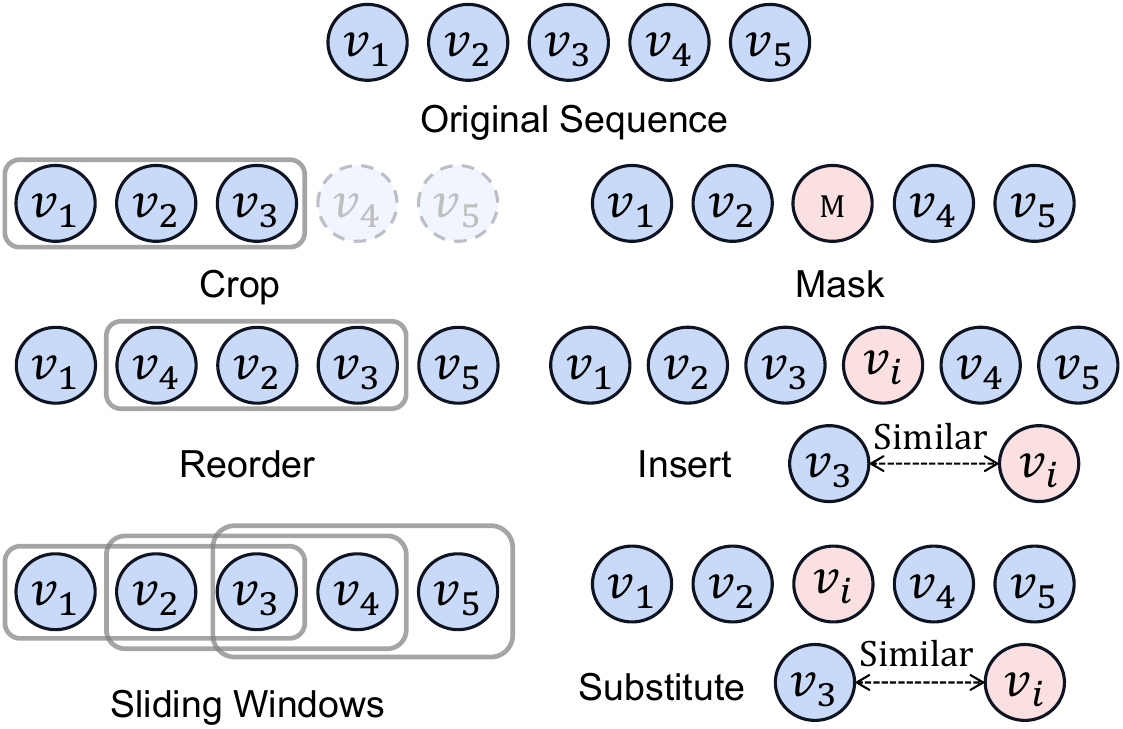}
  \vspace{-0.5em}
  \caption{Illustration of existing augmentation operators.}
  \label{fig:operators}
  \vspace{-1em}
\end{figure}

\subsection{Sequence Data Augmentation Operators}\label{sec:operators}
To alleviate the data sparsity, data augmentation is often used to augment the original data during training time. As illustrated in Figure \ref{fig:operators}, given an original sequence $s_u=[v_1, v_2, \ldots, v_{\left|s_u\right|}]$, we will introduce six of the most representative augmentation operators.

\begin{itemize}

\item \textbf{Crop:} Randomly select a continuous sub-sequence from $s_u$ \cite{xie2022contrastive}:
\begin{equation}
s_u^{a} = Aug_{crop}(s_u) = [v_i, v_{i+1}, \ldots, v_{i+L-1}],
\end{equation}
where $L$ is the length of sub-sequence.

\item \textbf{Reorder:} Randomly shuffle a continuous sub-sequence of the original sequence $s_u$ as $[v_i^{\prime}, \ldots, v_{i+r-1}^{\prime}]$ \cite{xie2022contrastive}:
\begin{equation}
s_u^{a} = Aug_{reorder}(s_u) = [v_1, v_2, \cdots, v_i^{\prime}, \cdots, v_{i+L-1}^{\prime}, \cdots, v_{\left|s_u\right|}].
\end{equation}

\item \textbf{Sliding Windows:} Given a window length $T$, this operation divides the original sequence into multiple sub-sequences by sliding a window from one end to the other \cite{tang2018personalized}:
\begin{equation}
\{s_u^{a_1},s_u^{a_2},\cdot\cdot\cdot,s_u^{a_n}\} = Aug_{windows}(s_u).
\end{equation}

\item \textbf{Mask:} Randomly mask a proportion of items in $s_u$ \cite{sun2019bert4rec,ren2023contrastive}:
\begin{equation}
s_u^{a} = Aug_{mask}(s_u) = [v_1^{\prime}, v_2^{\prime}, \ldots, v_{\left|s_u\right|}^{\prime}],
\end{equation}
where $v_i^{\prime}$ will be replaced with the `[mask]' token if $v_i$ is a selected item, otherwise $v_i^{\prime}=v_i$. 

\item \textbf{Substitute:} Replace a proportion of items in $s_u$ \cite{liu2021contrastive}:
\begin{equation}
s_u^{a} = Aug_{substitute}(s_u) = [v_1^{\prime}, v_2^{\prime}, \ldots, v_{\left|s_u\right|}^{\prime}],
\end{equation}
where $v_i^{\prime}$ will be replaced with the correlated item if $v_i$ is a selected item, otherwise $v_i^{\prime}=v_i$. The correlated item is obtained based on the correlation score or the similarity of item representation. 

\item \textbf{Insert:} Insert a proportion of items into $s_u$ \cite{liu2021contrastive}: 
\begin{equation}
s_u^{a} = Aug_{insert}(s_u) = [v_1, \ldots, v_{i}, v_{i}^{c}, \ldots, v_{\left|s_u\right|}],
\end{equation}
where $v_{i}^{c}$ is the item correlated to the insertion position. 

\end{itemize}

\subsection{Test-time Augmentation}
TTA augments the original inputs during the model inference to generate multiple variants. Afterward, the final result is obtained by average aggregating the model prediction on these variants \cite{kimura2021understanding}. Let $\boldsymbol{s}$ be the input sequence at test time. We can perform multiple data augmentations $Aug(\cdot)$ to $\boldsymbol{s}$ and get $\left\{\tilde{s}_i\right\}_{i=1}^m$, where $\tilde{\boldsymbol{s}}_i$ is the $i$-th augmented sequence and $m$ is the total number augmented sequences. Finally, the model predicts the most possible item $v^{*}$ based on the average aggregation of original input $\boldsymbol{s}$ as follows:
\begin{equation}\label{Eq:TTA}
v^{*}=AverageAgg\left(\sum_{i=1}^m Model\left(\tilde{\boldsymbol{s}}_i\right)\right). 
\end{equation}

TTA has been shown to be an effective and efficient augmentation method in many domains, such as computer vision \cite{kim2020learning, zhang2022memo} and graph learning \cite{ju2024graphpatcher, jin2022empowering}. In SR, although researchers have proposed many training-time augmentation methods, deploying them requires significant time and cost overhead. Meanwhile, whether existing operators can be used for TTA and whether simpler and more effective methods can be proposed is still unexplored.

\begin{table}[!t]
  \centering
  \caption{Results of empirical study with GRU4Rec.}
    \vspace{-1em}
  \renewcommand\arraystretch{0.95}
  \setlength{\tabcolsep}{1mm}{
  \scalebox{0.785}{
    \begin{tabular}{l|cccc|cccc}
    \toprule
     \multicolumn{1}{c|}{\multirow{2}[4]{*}{GRU4Rec}} & \multicolumn{4}{c|}{Beauty} & \multicolumn{4}{c}{Sports} \\
\cmidrule{2-9} & H@10 & N@10 & H@20 & N@20 & H@10 & N@10 & H@20 & N@20 \\
    \midrule
    Base & 0.0376 & 0.0180 & 0.0652 & 0.0249 & 0.0157 & 0.0075 & 0.0291 & 0.0108  \\
    \midrule
    + Mask & \underline{0.0435} & \textbf{0.0211} & \underline{0.0691} & \textbf{0.0284} & \underline{0.0241} & \underline{0.0121} & \underline{0.0416} & \underline{0.0164}  \\
    + Crop & 0.0387 & 0.0192 & 0.0623 & 0.0259 & 0.0232 & 0.0109 & 0.0392 & 0.0149  \\
    + Reorder & 0.0400 & 0.0194 & 0.0668 & 0.0262 & 0.0197 & 0.0093 & 0.0348 & 0.0131  \\
    + Substitute & \textbf{0.0439} & \underline{0.0209} & \textbf{0.0723} & \underline{0.0281} & \textbf{0.0252} & \textbf{0.0123} & \textbf{0.0426} & \textbf{0.0166} \\
    + Insert & 0.0394 & 0.0188 & 0.0679 & 0.0260 & 0.0206 & 0.0098 & 0.0361 & 0.0137  \\
    + Windows & 0.0359 & 0.0179 & 0.0577 & 0.0234 & 0.0226 & 0.0106 & 0.0389 & 0.0147  \\
    \midrule
    + CMR & 0.0398 & 0.0191 & 0.0660 & 0.0265 & 0.0215 & 0.0105 & 0.0371 & 0.0146  \\
    + CMRSI & 0.0413 & 0.0202 & 0.0683 & 0.0274 & 0.0228 & 0.0115 & 0.0396 & 0.0156  \\
    \bottomrule
    \end{tabular}}}%
  \label{tab:existing1}%
    \vspace{-1em}
\end{table}%

\begin{table}[!t]
  \centering
  \caption{Results of empirical study with SASRec.}
    \vspace{-1em}
  \renewcommand\arraystretch{0.95}
    \setlength{\tabcolsep}{1mm}{
  \scalebox{0.785}{
    \begin{tabular}{l|cccc|cccc}
    \toprule
     \multicolumn{1}{c|}{\multirow{2}[4]{*}{SASRec}} & \multicolumn{4}{c|}{Beauty} & \multicolumn{4}{c}{Sports} \\
\cmidrule{2-9} & H@10 & N@10 & H@20 & \multicolumn{1}{c|}{N@20} & H@10 & N@10 & H@20 & N@20 \\
    \midrule
    Base & 0.0584 & 0.0310 & 0.0914 & 0.0393 & 0.0317 & 0.0161 & 0.0495 & 0.0206  \\
    \midrule
    + Mask & \underline{0.0603} & \underline{0.0317} & \underline{0.0933} & \underline{0.0401} & \underline{0.0352} & \underline{0.0181} & \underline{0.0544} & \underline{0.0221}  \\
    + Crop & 0.0533 & 0.0282 & 0.0822 & 0.0354 & 0.0269 & 0.0138 & 0.0442 & 0.0182  \\
    + Reorder & 0.0584 & 0.0309 & 0.0913 & 0.0394 & 0.0320 & 0.0177 & 0.0494 & 0.0217  \\
    + Substitute & \textbf{0.0613} & \textbf{0.0329} & \textbf{0.0949} & \textbf{0.0413} & \textbf{0.0369} & \textbf{0.0195} & \textbf{0.0558} & \textbf{0.0243} \\
    + Insert & 0.0578 & 0.0307 & 0.0905 & 0.0390 & 0.0321 & 0.0176 & 0.0495 & 0.0220  \\
    + Windows & 0.0569 & 0.0304 & 0.0864 & 0.0378 & 0.0306 & 0.0154 & 0.0473 & 0.0196  \\
    \midrule
    + CMR & 0.0563 & 0.0295 & 0.0893 & 0.0383 & 0.0332 & 0.0162 & 0.0536 & 0.0208  \\
    + CMRSI & 0.0585 & 0.0308 & 0.0931 & 0.0406 & 0.0343 & 0.0175 & 0.0520 & 0.0215  \\
    \bottomrule
    \end{tabular}}}%
  \label{tab:existing2}%
    \vspace{-1em}
\end{table}%

\section{Empirical Study}

\subsection{Existing Operators For TTA (Q1)}\label{sec:existing}
We first explore the performance of existing operators in TTA. The experiments are conducted on two representative SR models, GRU4Rec \cite{hidasi2015session}, and SASRec \cite{kang2018self}, with two datasets, Amazon Beauty and Sports \cite{mcauley2015inferring}. We first fully train the model using the original dataset (without any augmentation), after which we add the operators introduced in Section \ref{sec:operators} For TTA. We average the probabilities of the output items. For the augmentation ratio of each operator, we carefully tune in the range of $[0.1,0.9]$ with steps of 0.1. Each input sequence during testing will be independently augmented ten times by the same operator, i.e., $m = 10$ in Eq \ref{Eq:TTA} (more details in Section \ref{Sec:setup}). The results are presented in Table \ref{tab:existing1} and \ref{tab:existing2}. Note that the CMR \cite{xie2022contrastive} is a Combination of Crop, Mask, and Reorder. CMRSI \cite{liu2021contrastive} adds Substitute and Insert to CMR. For combinations, each sequence is randomly augmented by one of the operators.

\vspace{0.3em}

\noindent \textbf{Observation.} Mask and Substitute perform better in all cases. Usually, the Substitute performs best, and the Mask performs second best. For the other operators, they only bring a slight performance gain, and in many cases, using them even brings an accuracy degradation. Combinations of multiple operators, CMR and CMRSI, do not perform as well as using only Substitute or Mask. The above results show the potential of existing operators to be used in TTA.

\vspace{0.3em}

\noindent \textbf{Hypothesis.} We hypothesize that the two better-performing operators produce appropriate perturbations on input sequences. These appropriate perturbations force the model to rely on more than just the learned preference patterns to make predictions, improving the model's generalization ability and prediction accuracy. Next, we will further validate and analyze our hypothesis.

\begin{table}[!t]
  \centering
  \caption{Similarity between augmented and original data.}
      \vspace{-1em}
    \renewcommand\arraystretch{0.9}
  \scalebox{0.85}{
    \begin{tabular}{c|cccc|c}
    \toprule
    \multirow{2}[4]{*}{Methods} & \multicolumn{2}{c}{SASRec} & \multicolumn{2}{c|}{GRU4Rec} & \multicolumn{1}{c}{\multirow{2}[4]{*}{Average}} \\
\cmidrule{2-5} & Beauty & Sports & Beauty & Sports & \\
    \midrule
    Mask & 0.9969 & 0.9772 & 0.9519 & 0.8745 & 0.9501 \\
    Substitute & 0.9786 & 0.9756 & 0.9232 & 0.8891 & 0.9416  \\
    \midrule
    Reorder & 0.9931 & 0.9958 & 0.8775 & 0.8094 & 0.9190  \\
    CMRSI & 0.9643 & 0.9549 & 0.8801 & 0.8378 & 0.9093  \\
    CMR & 0.9627 & 0.9411 & 0.8761 & 0.8045 & 0.8961  \\
    Insert & 0.9276 & 0.9169 & 0.7749 & 0.7606 & 0.8450  \\
    Windows & 0.9125 & 0.9086 & 0.7827 & 0.7349 & 0.8347  \\
    Crop & 0.9044 & 0.8993 & 0.7995 & 0.7305 & 0.8334  \\
    \bottomrule
    \end{tabular}}%
  \label{tab:cos}%
    \vspace{-1.5em}
\end{table}%

\subsection{What Makes Substitute and Mask Better (Q2)}
\subsubsection{Data Similarity and Sequential Pattern} In order to verify our hypothesis about appropriate perturbations, we calculated the cosine similarity between the augmented data obtained from different augmentation methods and the original data. Since ID-based sequence data cannot be computed directly, we compute the similarity of the output representation by the well-trained model. We report the average similarity throughout the inference process and present the result in Table \ref{tab:cos}. For comparison purposes, we rank the results from largest to smallest. 

\vspace{0.3em}

\noindent \textbf{Observation.} Combining Tables \ref{tab:existing1}, \ref{tab:existing2}, and \ref{tab:cos}, we can observe that the performance of the operator may correlated with the average similarity. The better-performing Substitute and Mask produced sequences with high similarity to the original sequence, while the poorer-performing Sliding Windows and Crop produced sequences with low similarity to the original data. Sliding Windows and Crop introduce too much perturbations into the original sequence. The above result validates our hypothesis of appropriate perturbations. 

\vspace{0.3em}

\noindent \textbf{Analysis from the operation.} Furthermore, as can be observed from Figure \ref{fig:operators}, both Insert and Reorder significantly impact the original sequential pattern when augmenting the original sequence. When inserting items into the sequence, all original items create new sequence relationships with the inserted items. When more than one item is inserted, the new sequence relationship becomes complex and different from the original sequence. The Reorder directly disrupts the original sequence order. Both Crop and Sliding Windows may intercept earlier subsequences of the original sequence and lose recent interactions, which is essential for predicting the user's next interactions \cite{chen2018sequential}. The augmented data produced by all four of the above operators have the potential to impact model predictions negatively. In contrast, the substitute replaces items with similar items, with minimal impact on the original sequential pattern. The mask allows the model to observe higher-order item relationships while preserving the global sequential patterns \cite{liu2021contrastive}.

\begin{figure}[!t]
  \centering
  \includegraphics[scale=0.52]{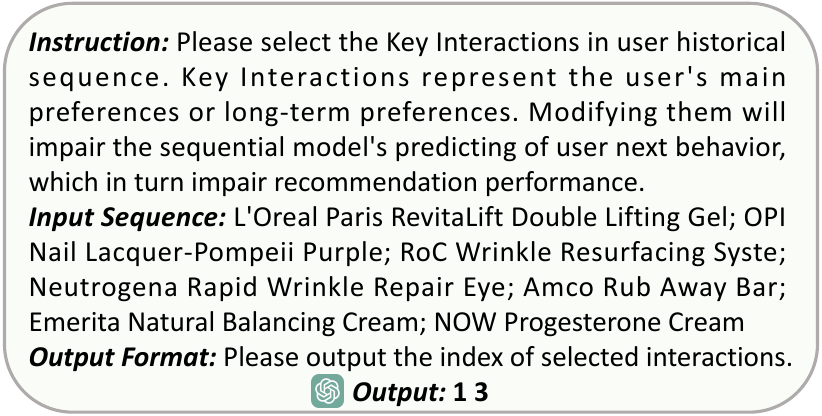}
  \vspace{-0.5em}
  \caption{Constructed prompt and example for LLMs to identify the key interactions based on user sequence.}
  \label{fig:prompt}
  \vspace{-1em}
\end{figure}

\begin{table}[!t]
  \centering
  \caption{The performance of different variants.}
    \vspace{-1em}
    \renewcommand\arraystretch{0.95}
        \setlength{\tabcolsep}{1mm}{
  \scalebox{0.775}{
    \begin{tabular}{l|cccc|cccc}
    \toprule
     \multicolumn{1}{c|}{\multirow{3}[3]{*}{Variants}} & \multicolumn{4}{c|}{Beauty} & \multicolumn{4}{c}{Sports} \\
\cmidrule{2-9}    \multicolumn{1}{c|}{} & \multicolumn{2}{c}{GRU4Rec} & \multicolumn{2}{c|}{SASRec} & \multicolumn{2}{c}{GRU4Rec} & \multicolumn{2}{c}{SASRec} \\
    \multicolumn{1}{c|}{} & \multicolumn{1}{c}{H@10} & \multicolumn{1}{c}{N@10} & \multicolumn{1}{c}{H@10} & \multicolumn{1}{c|}{N@10} & \multicolumn{1}{c}{H@10} & \multicolumn{1}{c}{N@10} & \multicolumn{1}{c}{H@10} & \multicolumn{1}{c}{N@10} \\
    \midrule
    Base  & 0.0376  & 0.0180  & 0.0584  & 0.0310  & 0.0157  & 0.0075  & 0.0317  & 0.0161  \\
    \midrule
    + Mask & 0.0435  & 0.0211  & 0.0603  & 0.0317  & 0.0241  & 0.0121  & 0.0352  & 0.0181  \\
    + Mask-KF & 0.0330  & 0.0140  & 0.0537  & 0.0289  & 0.0117  & 0.0057  & 0.0295  & 0.0142  \\
    + Mask-NKF & 0.0348  & 0.0168  & 0.0574  & 0.0305  & 0.0160  & 0.0077  & 0.0323  & 0.0154  \\
    + Mask-FR & 0.0384  & 0.0178  & 0.0597  & 0.0312  & 0.0210  & 0.0111  & 0.0332  & 0.0175  \\
    \midrule
    + Substitute & 0.0439  & 0.0209  & 0.0613  & 0.0329  & 0.0252  & 0.0123  & 0.0369  & 0.0195  \\
    + Substitute-KF & 0.0347  & 0.0167  & 0.0561  & 0.0304  & 0.0138  & 0.0068  & 0.0300  & 0.0153  \\
    + Substitute-NKF & 0.0389  & 0.0187  & 0.0593  & 0.0319  & 0.0189  & 0.0086  & 0.0326  & 0.0170  \\
    + Substitute-FR & 0.0420  & 0.0201  & 0.0604  & 0.0321  & 0.0227  & 0.0107  & 0.0346  & 0.0185  \\
    \bottomrule
    \end{tabular}}}%
  \label{tab:variant}%
\vspace{-1em}
\end{table}%

\subsubsection{Random Selection of Operating Interactions} When performing augmentation, existing operators randomly select a number of interactions in sequence for operation based on a given ratio. We explore the effect of random selection on TTA performance. Inspired by the recent research and capability of large language models (LLMs) \cite{zhang2024finerec,lin2023can,liu2024cora,Darec}, we utilize LLMs\footnote{In our experiments, we employ GPT-4 from \url{https://chat.openai.com/}.} to identify \emph{Key Interactions} in user sequences. Key Interactions represent the user's main preferences or long-term preferences. Modifying them will impair the model's predicting of user behavior, which impair the recommendation accuracy. We take the titles of the items in the original sequence as input to the LLMs. With a well-designed prompt and the powerful world knowledge of the LLMs, we can obtain the the key interactions in each sequence that best represent the user's preferences. This process can be formulated as follows: 
\begin{equation}
s_u^{k} = [v_1^k, v_2^k, \ldots, v_n^k] = LLMs(prompt, s_u),
\end{equation}
where $v_i^k \in s_u^{k}$ is the key interactions in $s_u$. The number of key interactions $n$ will be no more than half of the original sequence length. The non-key interactions $s_u^{nk}$ can be obtained as $s_u - s_u^{k}$. We give the prompt and specific example in Figure \ref{fig:prompt}. 

After obtaining the key interactions, we designed several variants of Mask and Substitute: 1) Key First (KF): Prioritize selecting operating interactions from $s_u^{k}$; 2) Non-Key First (NKF): Prioritize selecting operating interactions from $s_u^{nk}$; 3) Fixed Proportion (FR): Selecting from two sets in a fixed proportion. For this variant, we carefully tune the proportion in the range of $[0.05,0.95]$ with steps of 0.05. We present the result in Table \ref{tab:variant}. 

\vspace{0.3em}

\noindent \textbf{Observation.} The performance of the KF variant drops significantly, even below that of the base model. When NKF is used, it brings little improvement compared to the base model. This indicates that the LLMs successfully identify the key interactions. If we modify these interactions, it will have a negative impact on the model to predict the user's preference and behavior. Meanwhile, if we just modify the non-key interactions, TTA hardly brings further performance gains to the model. For variant FR, its performance is slightly lower than random selection (original operator). 

\vspace{0.3em}

\noindent \textbf{Analysis from the result.} Based on the above results, we believe that a appropriate choice when performing TTA on sequences should be a small number of key interactions and a large number of non-key interactions, both of which need to be satisfied simultaneously. Since the key are only a few in the sequence, the random selection of Mask and Substitute satisfies the above condition, achieving better performance.

\subsection{Summary and Limitations}\label{sec:summary}
Based on the above analysis, we summarize our findings as follows:
\begin{itemize}
    \item Existing operators can bring performance gains to the model when used for TTA, with Mask and Substitute being superior.
    \item Mask and Substitute introduce appropriate perturbations while preserving the original sequential patterns, which further improves the performance of the model during inference.
    \item The choice of augmenting location also has a crucial impact on the performance. The random selection satisfies the condition of a small number of key interactions and a large number of non-key interactions and achieves a better performance.
    \item Other operators, on the other hand, suffers from the problems of losing recent interactions, disrupting the original sequence pattern, or generating data that is too far from the original data.
\end{itemize}

\begin{figure}[!t]
  \centering
  \includegraphics[scale=0.40]{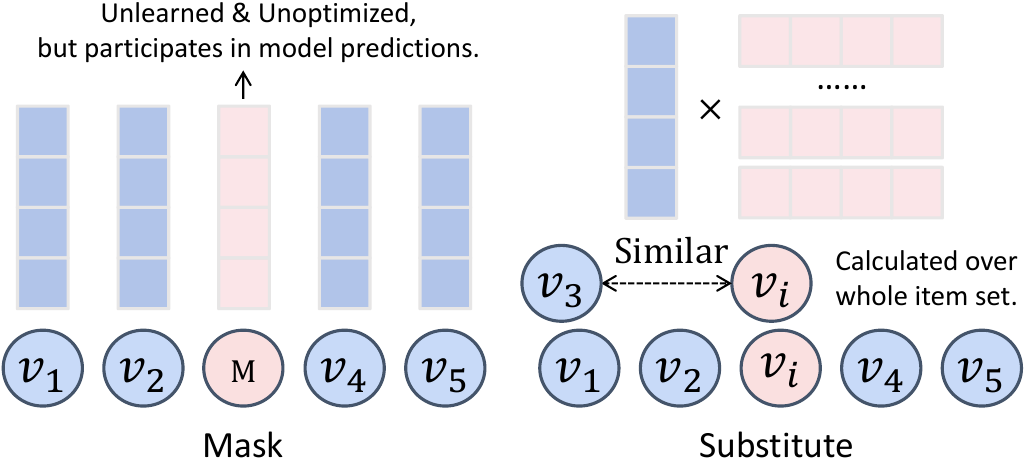}
  \vspace{-0.5em}
  \caption{Limitations of Mask and Substitute.}
  \label{fig:limitation}
  \vspace{-1em}
\end{figure}

\noindent \textbf{Limitations of Mask and Substitute.} So far, we have answered the first two questions. As shown in Figure \ref{fig:limitation}, we argue that the Substitute and Mask still have limitations. When performing the Substitute, the operator replaces the selected target item with the most similar item from the entire item set. For item $v_i$ and $v_j$. The similarity is usually calculated based on the dot product of their representations $e_i$, and $e_j$ \cite{liu2021contrastive}, i.e., $e_i \cdot e_j$. However, the computational and time overhead will grow significantly when the calculation is extended to the entire item set and to multiple substitution operations for a sequence. During inference, the computational complexity of this operation is $O(d|\mathcal{U}||\mathcal{V}|mn)$, where $d$, $|\mathcal{U}|$, $|\mathcal{V}|$, $m$, and $n$ are embedding dimension, number of users, number of items, number of times each sequence is augmented, and number of items replaced in each sequence, respectively. For Mask, the original operator uses a specific token for masking. However, following previous work \cite{liu2021contrastive,xie2022contrastive}, this token is often an extra special item in the Item set that no user has interacted with and contains no meaningful information. In other words, its item embedding is not learned and optimized (initialized only), especially when we use Mask directly in TTA. During inference, these unlearned and unoptimized embeddings still participate in the model calculation, which interfere the model prediction and thus impair the performance.

\section{Our Method}

\subsection{TNoise and TMask (Q3)}

\begin{figure}[!t]
  \centering
  \includegraphics[scale=0.40]{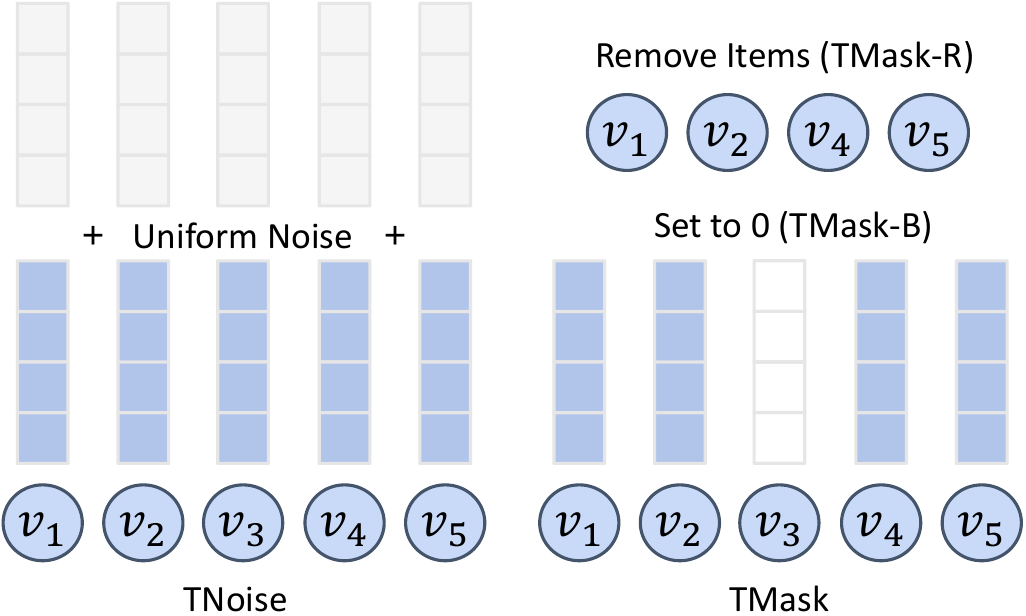}
  \vspace{-0.5em}
  \caption{Illustration of our proposed TNoise and TMask.}
  \label{fig:method}
  \vspace{-1em}
\end{figure}

Previous analyses have shown that the key to TTA is appropriate perturbations with no destruction of the sequential patterns. In addition, the location of the operation needs to satisfy randomness. Based on these analyses, we present our TNoise and TMask. We give an illustration of our method in Figure \ref{fig:method}.

\vspace{0.3em}

\noindent \textbf{TNoise.} Inspired by the previous work in graph contrastive learning and perturbation on images \cite{goodfellow2014explaining,yu2022graph}, we propose directly adding uniform noises to the sequence representation for efficient and effective augmentation. We call this operator TNoise. Give an original sequence $s_u=[v_1, \ldots, v_j, \ldots, v_{\left|s_u\right|}]$, we usually applied the $\operatorname{Look-Up}$ from item embedding matrix $\mathbb{R}^{|\mathcal{V}| \times d}$ to get a sequence of item representation, i.e., $E_u=[e_{v_1}, e_{v_2}, \ldots, e_{v_{\left|s_u\right|}}]$. The item embedding matrix projects the high-dimensional one-hot representation of an item to low-dimensional dense representations. After that, the TNoise can be formulated as follows:
\begin{equation}\label{eq:noise}
E_u^\prime = Aug_{TNoise}(E_u)= E_u + \epsilon \text{, where}\; \epsilon \in \mathcal{U}\left(a, b\right),
\end{equation}
where $a$ and $b$ are hyperparameters to control the noise variance. In addition, this operation can be performed on the hidden state of the sequence. By feeding the sequence representation $E_u$ into the $\operatorname{Encoder}$, we can obtain the hidden state of the sequence $H_u$. The $H_u$ can also be augmented by Eq.\ref{eq:noise} and get $H_u^\prime$. TNoise avoids the computational overhead associated with Substitute while satisfying the condition mentioned above. It does not disrupt the original sequential patterns, while by adjusting $a$ and $b$, we can introduce appropriate perturbations. Typically, better performance is obtained by setting $a$ and $b$ to 1 and 0.5, respectively. During inference, its complexity is $O(d|\mathcal{U}|m)$, significantly lower than $O(d|\mathcal{U}||\mathcal{V}|mn)$.

\vspace{0.3em}

\noindent \textbf{TMask.} When using the original Mask, the model is negatively affected by the representation of unlearned and unoptimized Mask tokens. Therefore, we propose to improve the Mask by blocking the mask tokens from participating in model calculations. To achieve that, we set all the item embeddings corresponding to the mask tokens' location to 0 after the $\operatorname{Look-Up}$ operation. We call this variant TMask-B, which can be formulated as follows: 
\begin{equation}
E_u^\prime = Aug_{TMask-B}(E_u) = [e_{v_1}^{\prime}, e_{v_2}^{\prime}, \ldots, e_{v_{\left|s_u\right|}}^{\prime}],
\end{equation}
where $e_{v_i}^{\prime}$ will be set to zero embedding if $e_{v_i}$ is selected, otherwise $e_{v_i}^{\prime}=e_{v_i}$. Further, we present another variant, called TMask-R, by directly removing the items that should have become mask tokens. The TMask-R can be formulated as follows:
\begin{equation}
s_u^{a} = Aug_{TMask-R}(s_u) = [v_1, v_2, \ldots, v_{\left|s_u^{a}\right|}],
\end{equation}
where $s_u^{a}$ is the sequence after deleting. Following our previous analysis, the operation locations of the TMask are all randomly selected. Given the sequence length $L$, the number of operated items in each sequence is $L\sigma$, where $\sigma \in (0,1)$ is a hyperparameter.

\subsection{Discussions}
\noindent \textbf{Advantages of Our Method.} Compared to existing augmentation methods, our approach has the following advantages:

\begin{itemize}
    \item \textbf{Satisfies all conditions:} TNoise and the two variants of TMask satisfy all the conditions we derived in Section \ref{sec:summary}. They introduce appropriate perturbations while preserving the original sequential patterns. Randomized location selection results in better performance without the consideration for other information or additional calculations when selecting a location \cite{dang2023uniform,dang2023ticoserec}. TMask makes high-order item relationships available during inference.
    \item \textbf{No training required and low time cost:} Since the augmentation is performed at test time, there is no need to training the model when deploying, which will save computational and time overheads. During inference, the time complexity of both operators is $O(d|\mathcal{U}|m)$, which is lower than the Substitute.
    \item \textbf{Model agnostic and no learnable parameters:} Some researchers use bidirectional Transformers \cite{liu2021augmenting} or diffusion models \cite{liu2023diffusion} for augmentation. These methods have limitations on the type of backbone network and usually introduce additional model parameters. In contrast, our method does not contain any learnable parameters and can be adapted to most sequential models. 
\end{itemize}

% \subsubsection{Possible Limitations.} 

\noindent \textbf{Possible Limitations.} Although empirical studies have shown the effectiveness of TTA, the current improvement that TTA can provide may be lower than training-time augmentation. The latter provides more data during training and can help the model learn user preferences and sequential patterns better. Better recommendation performance may achieved by combining both training-time and test-time augmentations. In addition, several studies \cite{shanmugam2020and,kimura2021understanding} have shown that TTA may fail, i.e., not lead to performance improvement, when a specific model is combined with a specific TTA method. This phenomenon is related to model architectures, datasets, and augmentation types. We leave this for future research.

\begin{table*}[!t]
  \centering
  \caption{Performance comparison of different test-time augmentation methods. The best performance is bolded and the second best is underlined. The `Inf. Time' represents Inference Time (minutes). The `Improve-$\alpha$' and `Improve-$\beta$' represent the relative improvements of our best result over the based model and the best baseline method, respectively.}
  \vspace{-1em}
    \renewcommand\arraystretch{0.9}
    \setlength{\tabcolsep}{1.5mm}{
      \scalebox{0.85}{
    \begin{tabular}{l|ccccccc|ccccccc}
    \toprule
    \multicolumn{1}{c|}{\multirow{2}[1]{*}{Method}} & \multicolumn{7}{c|}{Beauty} & \multicolumn{7}{c}{Sports} \\
    \multicolumn{1}{c|}{} & \multicolumn{1}{c}{H@5} & \multicolumn{1}{c}{N@5} & H@10 & N@10 & H@20 & N@20 & \multicolumn{1}{c|}{Inf. Time} & 
    \multicolumn{1}{c}{H@5} & \multicolumn{1}{c}{N@5} & H@10 & N@10 & H@20 & N@20 & \multicolumn{1}{c}{inf. Time} \\
    \midrule
    GRU4Rec (Base) & 0.0204 & 0.0125 & 0.0376 & 0.0180 & 0.0652 & 0.0249 & \multicolumn{1}{c|}{0.1} & 0.0083 & 0.0051 & 0.0157 & 0.0075 & 0.0291 & 0.0108 & \multicolumn{1}{c}{0.2} \\
    \midrule
    + Mask & 0.0241 & \underline{0.0152} & 0.0435 & \underline{0.0211} & 0.0691 & \underline{0.0284} & \multicolumn{1}{c|}{0.1} & \underline{0.0142} & \underline{0.0089} & 0.0241 & 0.0121 & 0.0416 & 0.0164 & \multicolumn{1}{c}{0.3} \\
    + Substitute & \underline{0.0247} & 0.0149 & \underline{0.0439} & 0.0209 & \underline{0.0723} & 0.0281 & \multicolumn{1}{c|}{0.6} & 0.0141 & 0.0087 & \underline{0.0252} & \underline{0.0123} & \underline{0.0426} & \underline{0.0166} & \multicolumn{1}{c}{0.9} \\
    + CMR & 0.0229 & 0.0147 & 0.0398 & 0.0191 & 0.0660 & 0.0265 & \multicolumn{1}{c|}{0.1} & 0.0123 & 0.0075 & 0.0215 & 0.0105 & 0.0371 & 0.0146 & \multicolumn{1}{c}{0.4} \\
    + CMRSI & 0.0234 & 0.0149 & 0.0413 & 0.0202 & 0.0683 & 0.0274 & \multicolumn{1}{c|}{0.6} & 0.0136 & 0.0083 & 0.0228 & 0.0115 & 0.0396 & 0.0156 & \multicolumn{1}{c}{0.8} \\
    \midrule
    + TNoise (Ours) & 0.0209 & 0.0137 & 0.0375 & 0.0181 & 0.0647 & 0.0252 & \multicolumn{1}{c|}{0.1} & 0.0080 & 0.0050 & 0.0156 & 0.0074 & 0.0290 & 0.0107 & \multicolumn{1}{c}{0.3} \\
    + TMask-B (Ours) & 0.0224 & 0.0149 & 0.0395 & 0.0196 & 0.0670 & 0.0269 & \multicolumn{1}{c|}{0.1} & 0.0137 & 0.0088 & 0.0236 & 0.0120 & 0.0430 & 0.0161 & \multicolumn{1}{c}{0.3} \\
    + TMask-R (Ours) & \textbf{0.0289} & \textbf{0.0181} & \textbf{0.0463} & \textbf{0.0237} & \textbf{0.0739} & \textbf{0.0307} & \multicolumn{1}{c|}{0.1} & \textbf{0.0162} & \textbf{0.0102} & \textbf{0.0274} & \textbf{0.0138} & \textbf{0.0468} & \textbf{0.0187} & \multicolumn{1}{c}{0.3} \\
    Improve-$\alpha$ & 41.67\% & 44.80\% & 23.14\% & 31.67\% & 13.34\% & 23.29\% & - & 95.18\% & 100.00\% & 74.52\% & 84.00\% & 60.82\% & 73.15\% & - \\
    Improve-$\beta$ & 17.00\% & 19.08\% & 5.47\% & 12.32\% & 2.21\% & 8.10\% & - & 14.08\% & 14.61\% & 8.73\% & 12.20\% & 9.86\% & 12.65\% & - \\
    \midrule
    \midrule
    SASRec (Base) & 0.0373 & 0.0243 & 0.0584 & 0.0310 & 0.0914 & 0.0393 & \multicolumn{1}{c|}{0.1} & 0.0190 & 0.0121 & 0.0317 & 0.0161 & 0.0495 & 0.0206 & \multicolumn{1}{c}{0.4} \\
    \midrule
    + Mask & 0.0382 & 0.0248 & 0.0603 & 0.0317 & 0.0933 & 0.0401 & \multicolumn{1}{c|}{0.1} & 0.0209 & 0.0140 & 0.0352 & 0.0181 & 0.0544 & 0.0221 & \multicolumn{1}{c}{0.5} \\
    + Substitute & 0.0396 & 0.0259 & 0.0613 & 0.0329 & 0.0949 & 0.0413 & \multicolumn{1}{c|}{0.7} & 0.0224 & 0.0149 & 0.0369 & 0.0195 & 0.0558 & 0.0243 & \multicolumn{1}{c}{1.4} \\
    + CMR & 0.0360 & 0.0230 & 0.0563 & 0.0295 & 0.0893 & 0.0383 & \multicolumn{1}{c|}{0.1} & 0.0197 & 0.0119 & 0.0332 & 0.0162 & 0.0536 & 0.0208 & \multicolumn{1}{c}{0.5} \\
    + CMRSI & 0.0375 & 0.0243 & 0.0585 & 0.0308 & 0.0931 & 0.0406 & \multicolumn{1}{c|}{0.7} & 0.0209 & 0.0132 & 0.0343 & 0.0175 & 0.0520 & 0.0215 & \multicolumn{1}{c}{1.2} \\
    \midrule
    + TNoise (Ours) & \underline{0.0415} & \underline{0.0268} & \textbf{0.0653} & \textbf{0.0345} & \textbf{0.1001} & \underline{0.0432} & \multicolumn{1}{c|}{0.1} & \underline{0.0261} & \underline{0.0161} & \underline{0.0422} & \underline{0.0213} & \underline{0.0644} & \underline{0.0269} & \multicolumn{1}{c}{0.5} \\
    + TMask-B (Ours) & 0.0367 & 0.0237 & 0.0584 & 0.0307 & 0.0935 & 0.0395 & \multicolumn{1}{c|}{0.1} & 0.0213 & 0.0137 & 0.0355 & 0.0179 & 0.0542 & 0.0218 & \multicolumn{1}{c}{0.5} \\
    + TMask-R (Ours) & \textbf{0.0422} & \textbf{0.0272} & \underline{0.0646} & \underline{0.0344} & \textbf{0.1001} & \textbf{0.0433} & \multicolumn{1}{c|}{0.1} & \textbf{0.0267} & \textbf{0.0172} & \textbf{0.0432} & \textbf{0.0225} & \textbf{0.0667} & \textbf{0.0284} & \multicolumn{1}{c}{0.5} \\
    Improve-$\alpha$ & 13.14\% & 11.93\% & 11.82\% & 11.29\% & 9.52\% & 10.18\% & - & 40.53\% & 42.15\% & 36.28\% & 39.75\% & 34.75\% & 37.86\% & - \\
    Improve-$\beta$ & 6.57\% & 5.02\% & 6.53\% & 4.86\% & 5.48\% & 4.84\% & - & 19.20\% & 15.44\% & 17.07\% & 15.38\% & 19.53\% & 16.87\% & - \\
    \midrule
    \multicolumn{1}{r}{} &  &  &  &  &  &  & \multicolumn{1}{r}{} &  &  &  &  &  &   & \multicolumn{1}{r}{} \\
    \midrule
    \multicolumn{1}{c|}{\multirow{2}[1]{*}{Method}} & \multicolumn{7}{c|}{Home} & \multicolumn{7}{c}{Yelp} \\
    \multicolumn{1}{c|}{} & \multicolumn{1}{c}{H@5} & \multicolumn{1}{c}{N@5} & H@10 & N@10 & H@20 & N@20 & \multicolumn{1}{c|}{Inf. Time} & \multicolumn{1}{c}{H@5} & \multicolumn{1}{c}{N@5} & H@10 & N@10 & H@20 & N@20 & \multicolumn{1}{c}{Inf. Time} \\
    \midrule
    GRU4Rec (Base) & 0.0035 & 0.0022 & 0.0063 & 0.0031 & 0.0116 & 0.0044 & \multicolumn{1}{c|}{0.5} & 0.0082 & 0.0048 & 0.0158 & 0.0072 & 0.0298 & 0.0108 & \multicolumn{1}{c}{5.4} \\
    \midrule
    + Mask & 0.0045 & 0.0026 & 0.0085 & 0.0040 & 0.0161 & 0.0057 & \multicolumn{1}{c|}{0.6} & \underline{0.0101} & 0.0058 & 0.0183 & 0.0082 & \underline{0.0336} & \underline{0.0124} & \multicolumn{1}{c}{5.7} \\
    + Substitute & \underline{0.0049} & \underline{0.0032} & \underline{0.0094} & \underline{0.0044} & \underline{0.0178} & \underline{0.0065} & \multicolumn{1}{c|}{2.6} & 0.0097 & 0.0056 & \underline{0.0186} & \underline{0.0086} & 0.0331 & 0.0120 & \multicolumn{1}{c}{76.5} \\
    + CMR & 0.0038 & 0.0025 & 0.0068 & 0.0034 & 0.0142 & 0.0049 & \multicolumn{1}{c|}{0.7} & 0.0096 & \underline{0.0059} & 0.0179 & 0.0078 & 0.0323 & 0.0118 & \multicolumn{1}{c}{5.8} \\
    + CMRSI & 0.0046 & 0.0028 & 0.0084 & 0.0041 & 0.0170 & 0.0062 & \multicolumn{1}{c|}{2.5} & 0.0092 & 0.0053 & 0.0175 & 0.0079 & 0.0329 & 0.0118 & \multicolumn{1}{c}{44.6} \\
    \midrule
    + TNoise (Ours) & 0.0036 & 0.0021 & 0.0063 & 0.0032 & 0.0116 & 0.0043 & \multicolumn{1}{c|}{0.6} & 0.0081 & 0.0047 & 0.0157 & 0.0072 & 0.0296 & 0.0108 & \multicolumn{1}{c}{5.7} \\
    + TMask-B (Ours) & \underline{0.0055} & \underline{0.0035} & \underline{0.0098} & \underline{0.0048} & \underline{0.0172} & \underline{0.0067} & \multicolumn{1}{c|}{0.6} & 0.0093 & 0.0057 & 0.0180 & 0.0081 & 0.0322 & 0.0117 & \multicolumn{1}{c}{5.6} \\
    + TMask-R (Ours) & \textbf{0.0060} & \textbf{0.0037} & \textbf{0.0109} & \textbf{0.0053} & \textbf{0.0204} & \textbf{0.0076} & \multicolumn{1}{c|}{0.6} & \textbf{0.0114} & \textbf{0.0069} & \textbf{0.0207} & \textbf{0.0099} & \textbf{0.0365} & \textbf{0.0139} & \multicolumn{1}{c}{5.6} \\
    Improve-$\alpha$ & 71.43\% & 68.18\% & 73.02\% & 70.97\% & 75.86\% & 72.73\% & - & 39.02\% & 43.75\% & 31.01\% & 37.50\% & 22.48\% & 28.70\% & - \\
    Improve-$\beta$ & 22.45\% & 15.63\% & 15.96\% & 20.45\% & 14.61\% & 16.92\% & - & 12.87\% & 16.95\% & 11.29\% & 15.12\% & 8.63\% & 12.10\% & - \\
    \midrule
    \midrule
    SASRec (Base) & 0.0097 & 0.0061 & 0.0160 & 0.0081 & 0.0249 & 0.0104 & \multicolumn{1}{c|}{0.5} & 0.0142 & 0.0088 & 0.0253 & 0.0124 & 0.0430 & 0.0168 & \multicolumn{1}{c}{4.5} \\
    \midrule
    + Mask & 0.0101 & 0.0064 & 0.0165 & 0.0084 & 0.0257 & 0.0107 & \multicolumn{1}{c|}{0.6} & 0.0145 & 0.0095 & 0.0268 & 0.0131 & 0.0456 & 0.0178 & \multicolumn{1}{c}{6.3} \\
    + Substitute & 0.0112 & 0.0075 & 0.0186 & 0.0096 & 0.0278 & 0.0126 & \multicolumn{1}{c|}{3.7} & 0.0149 & 0.0097 & 0.0279 & 0.0136 & 0.0475 & 0.0183 & \multicolumn{1}{c}{113.1} \\
    + CMR & 0.0104 & 0.0063 & 0.0167 & 0.0085 & 0.0254 & 0.0106 & \multicolumn{1}{c|}{0.6} & 0.0143 & 0.0090 & 0.0256 & 0.0127 & 0.0435 & 0.0171 & \multicolumn{1}{c}{6.4} \\
    + CMRSI & 0.0106 & 0.0069 & 0.0175 & 0.0090 & 0.0263 & 0.0114 & \multicolumn{1}{c|}{3.3} & 0.0152 & 0.0099 & 0.0274 & 0.0134 & 0.0471 & 0.0176 & \multicolumn{1}{c}{109.8} \\
    \midrule
    + TNoise (Ours) & \textbf{0.0119} & \underline{0.0075} & \underline{0.0195} & \underline{0.0099} & \underline{0.0314} & \underline{0.0129} & \multicolumn{1}{c|}{0.6} & \textbf{0.0171} & \textbf{0.0108} & \underline{0.0293} & \textbf{0.0147} & \underline{0.0496} & \textbf{0.0198} & \multicolumn{1}{c}{6.1} \\
    + TMask-B (Ours) & 0.0098 & 0.0062 & 0.0161 & 0.0082 & 0.0261 & 0.0107 & \multicolumn{1}{c|}{0.6} & 0.0151 & 0.0093 & 0.0264 & 0.0129 & 0.0450 & 0.0176 & \multicolumn{1}{c}{6.0} \\
    + TMask-R (Ours) & \underline{0.0117} & \textbf{0.0078} & \textbf{0.0197} & \textbf{0.0101} & \textbf{0.0317} & \textbf{0.0131} & \multicolumn{1}{c|}{0.6} & \underline{0.0170} & \underline{0.0105} & \textbf{0.0299} & \underline{0.0146} & \textbf{0.0505} & \underline{0.0196} & \multicolumn{1}{c}{6.4} \\
    Improve-$\alpha$ & 22.68\% & 27.87\% & 23.13\% & 24.69\% & 27.31\% & 25.96\% & - & 20.42\% & 22.73\% & 18.18\% & 18.55\% & 17.44\% & 17.86\% & - \\
    Improve-$\beta$ & 6.25\% & 4.00\% & 5.91\% & 5.21\% & 14.03\% & 3.97\% & - & 12.50\% & 9.09\% & 7.17\% & 8.09\% & 6.32\% & 8.20\% & - \\
    \bottomrule
    \end{tabular}}}%
  \label{tab:main_result_1}%
    \vspace{-1em}
\end{table*}%

\begin{table}[!t]
	\centering
      \caption{The statistics of four datasets.}
         \vspace{-1em}
	\scalebox{0.9}{
		\begin{tabular}{l|rrrr}
			\toprule \textbf{{Dataset}} & \textbf{Beauty} &\textbf{Sports} & \textbf{Home} & \textbf{Yelp} \\
			\midrule 
		   {\# Users} & 22,363 & 35,958 & 66,519 & 213,170 \\
			 {\# Items} & 12,101 & 18,357 & 28,237 & 94,304 \\
			 {\# Interactions} & 198,502 & 296,337 & 551,682 & 3,277,392 \\
              {\# Average Length} & 8.9 & 8.3 & 8.3 & 15.4 \\
			 {Sparsity} & 99.92\% & 99.95\% & 99.97\% & 99.98\% \\
			\bottomrule
	\end{tabular}}
	\label{tab:datasets}
         \vspace{-1em}
\end{table}

\section{Experiments}
\subsection{Experiment Setup}\label{Sec:setup}
\subsubsection{Datasets} The experiments are conducted on four datasets. Beauty, Sports, and Home are obtained from Amazon \cite{mcauley2015inferring} and correspond to the "Beauty", "Sports and Outdoors", and "Home" categories, respectively. Yelp\footnote{\url{https://www.yelp.com/dataset}}is a business dataset, which we use its first version released in 2018. We reduce the data by extracting the 5-core \cite{liu2021contrastive,dang2023ticoserec}. The maximum sequence length is set to 50 for all datasets. The detailed statistics are summarized in Table \ref{tab:datasets}.

\begin{table*}[!t]
  \centering
  \caption{Performance comparison of our proposed test-time augmentation method with training-time augmentation methods. The best performance is bolded and the second best is underlined. The `Ret. Time' and `Inf. Time' represent the Retraining Time when deploying the methods and Inference Time, respectively (minutes).}
    \vspace{-1em}
    \renewcommand\arraystretch{0.95}
    \setlength{\tabcolsep}{1mm}{
      \scalebox{0.8}{
    \begin{tabular}{c|l|cccccccc|cccccccc}
    \toprule
    \multicolumn{1}{c|}{\multirow{2}[1]{*}{Aug. Stage}} & \multicolumn{1}{c|}{\multirow{2}[1]{*}{Method}} & \multicolumn{8}{c|}{Beauty}                        & \multicolumn{8}{c}{Sports} \\
    \multicolumn{1}{c|}{} & \multicolumn{1}{c|}{} & \multicolumn{1}{c}{H@5} & \multicolumn{1}{c}{N@5} & \multicolumn{1}{c}{H@10} & \multicolumn{1}{c}{N@10} & \multicolumn{1}{c}{H@20} & \multicolumn{1}{c}{N@20} & Ret. Time & \multicolumn{1}{c|}{Inf. Time} & \multicolumn{1}{c}{H@5} & \multicolumn{1}{c}{N@5} & \multicolumn{1}{c}{H@10} & \multicolumn{1}{c}{N@10} & \multicolumn{1}{c}{H@20} & \multicolumn{1}{c}{N@20} & Ret. Time & \multicolumn{1}{c}{Inf. Time} \\
    \midrule
    \multicolumn{2}{c|}{GRU4Rec (Base)} & 0.0204 & 0.0125 & 0.0376 & 0.0180 & 0.0652 & 0.0249 & -    & 0.1  & 0.0083 & 0.0051 & 0.0157 & 0.0075 & 0.0291 & 0.0108 & -    & 0.2  \\
        \midrule
    \multicolumn{1}{c|}{\multirow{4}[0]{*}{Training}} & + Windows & 0.0238 & 0.0143 & 0.0409 & 0.0203 & 0.0692 & 0.0271 & \multicolumn{1}{c}{23.4} & 0.1  & 0.0094 & 0.0055 & 0.0188 & 0.0085 & 0.0354 & 0.0127 & \multicolumn{1}{c}{31.3} & 0.2  \\
         & + CMR & 0.0268 & 0.0172 & 0.0446 & 0.0223 & 0.0705 & 0.0284 & \multicolumn{1}{c}{23.3} & 0.1  & 0.0118 & 0.0070 & 0.0219 & 0.0103 & 0.0392 & 0.0139 & \multicolumn{1}{c}{29.8} & 0.2  \\
         & + CMRSI & 0.0279 & 0.0177 & 0.0459 & 0.0234 & 0.0732 & 0.0299 & \multicolumn{1}{c}{242.5} & 0.1  & 0.0127 & 0.0076 & 0.0224 & 0.0107 & 0.0418 & 0.0156 & \multicolumn{1}{c}{372.3} & 0.2  \\
         & + CL4SRec & \textbf{0.0413} & \textbf{0.0269} & \textbf{0.0634} & \textbf{0.0340} & \textbf{0.0966} & \textbf{0.0423} & \multicolumn{1}{c}{26.3} & 0.1  & \textbf{0.0229} & \textbf{0.0145} & \textbf{0.0373} & \textbf{0.0191} & \textbf{0.0597} & \textbf{0.0247} & \multicolumn{1}{c}{58.3} & 0.2  \\
        \midrule
    \multicolumn{1}{c|}{Testing} & + TMask-R & \underline{0.0289} & \underline{0.0181} & \underline{0.0463} & \underline{0.0237} & \underline{0.0739} & \underline{0.0307} & -    & 0.1  & \underline{0.0162} & \underline{0.0102} & \underline{0.0274} & \underline{0.0138} & \underline{0.0468} & \underline{0.0187} & -    & 0.3  \\
    \midrule
    \midrule
    \multicolumn{2}{c|}{SASRec (Base)} & 0.0373 & 0.0243 & 0.0584 & 0.0310 & 0.0914 & 0.0393 & -    & 0.1  & 0.0190 & 0.0121 & 0.0317 & 0.0161 & 0.0495 & 0.0206 & -    & 0.4  \\
    \midrule
    \multicolumn{1}{c|}{\multirow{4}[2]{*}{Training}} & + Windows & 0.0365 & 0.0237 & 0.0591 & 0.0309 & 0.0938 & 0.0396 & \multicolumn{1}{c}{35.4} & 0.1  & 0.0225 & 0.0148 & 0.0359 & 0.0187 & 0.0544 & 0.0243 & \multicolumn{1}{c}{42.9} & 0.4  \\
         & + CMR & 0.0390 & 0.0253 & 0.0589 & 0.0317 & 0.0907 & 0.0396 & \multicolumn{1}{c}{23.8} & 0.1  & 0.0233 & 0.0159 & 0.0396 & 0.0203 & 0.0602 & 0.0252 & \multicolumn{1}{c}{37.6} & 0.4  \\
         & + CMRSI & 0.0401 & 0.0262 & 0.0619 & 0.0330 & 0.0952 & 0.0411 & \multicolumn{1}{c}{57.2} & 0.1  & 0.0250 & 0.0168 & 0.0393 & 0.0207 & 0.0606 & 0.0265 & \multicolumn{1}{c}{76.3} & 0.4  \\
         & + CL4SRec & \textbf{0.0477} & \textbf{0.0321} & \textbf{0.0721} & \textbf{0.0399} & \textbf{0.1048} & \textbf{0.0481} & \multicolumn{1}{c}{37.4} & 0.1  & \textbf{0.0300} & \textbf{0.0196} & \textbf{0.0474} & \textbf{0.0252} & \textbf{0.0699} & \textbf{0.0308} & \multicolumn{1}{c}{39.0} & 0.4  \\
    \midrule
    \multicolumn{1}{c|}{\multirow{2}[2]{*}{Testing}} & + TNoise & 0.0415 & 0.0268 & \underline{0.0653} & \underline{0.0345} & \underline{0.1001} & 0.0432 & -    & 0.1  & 0.0261 & 0.0161 & 0.0422 & 0.0213 & 0.0644 & 0.0269 & -    & 0.5  \\
         & + TMask-R & \underline{0.0422} & \underline{0.0272} & 0.0646 & 0.0344 & \underline{0.1001} & \underline{0.0433} & -    & 0.1  & \underline{0.0267} & \underline{0.0172} & \underline{0.0432} & \underline{0.0225} & \underline{0.0667} & \underline{0.0284} & -    & 0.5  \\
    \midrule
    \multicolumn{1}{c}{} & \multicolumn{1}{r}{} &      &      &      &      &      &      & \multicolumn{1}{r}{} & \multicolumn{1}{r}{} &      &      &      &      &      &      & \multicolumn{1}{r}{} &  \\
    \midrule
    \multicolumn{1}{c|}{\multirow{2}[2]{*}{Aug. Stage}} & \multicolumn{1}{c|}{\multirow{2}[1]{*}{Method}} & \multicolumn{8}{c|}{Home}                          & \multicolumn{8}{c}{Yelp} \\
    \multicolumn{1}{c|}{} & \multicolumn{1}{c|}{} & \multicolumn{1}{c}{H@5} & \multicolumn{1}{c}{N@5} & \multicolumn{1}{c}{H@10} & \multicolumn{1}{c}{N@10} & \multicolumn{1}{c}{H@20} & \multicolumn{1}{c}{N@20} & Ret. Time & \multicolumn{1}{c|}{Inf. Time} & \multicolumn{1}{c}{H@5} & \multicolumn{1}{c}{N@5} & \multicolumn{1}{c}{H@10} & \multicolumn{1}{c}{N@10} & \multicolumn{1}{c}{H@20} & \multicolumn{1}{c}{N@20} & Ret. Time & \multicolumn{1}{c}{Inf. Time} \\
    \midrule
    \multicolumn{2}{c|}{GRU4Rec (Base)} & 0.0035 & 0.0022 & 0.0063 & 0.0031 & 0.0116 & 0.0044 & -    & 0.5  & 0.0082 & 0.0048 & 0.0158 & 0.0072 & 0.0298 & 0.0108 & -    & 5.4  \\
    \midrule
    \multicolumn{1}{c|}{\multirow{4}[1]{*}{Training}} & + Windows & 0.0041 & 0.0025 & 0.0075 & 0.0036 & 0.0144 & 0.0053 & \multicolumn{1}{c}{58.9} & 0.5  & \underline{0.0142} & \underline{0.0091} & \underline{0.0249} & 0.0120 & 0.0429 & 0.0168 & \multicolumn{1}{c}{401.4} & 5.4  \\
         & + CMR & 0.0050 & 0.0031 & 0.0087 & 0.0042 & 0.0167 & 0.0061 & \multicolumn{1}{c}{43.4} & 0.5  & 0.0128 & 0.0082 & 0.0232 & 0.0115 & 0.0393 & 0.0157 & \multicolumn{1}{c}{308.9} & 5.4  \\
         & + CMRSI & 0.0057 & 0.0035 & 0.0105 & 0.0050 & 0.0177 & 0.0068 & \multicolumn{1}{c}{644.8} & 0.5  & 0.0139 & 0.0087 & 0.0243 & \underline{0.0126} & \underline{0.0441} & \underline{0.0170} & \multicolumn{1}{c}{5207.3} & 5.4  \\
         & + CL4SRec & \textbf{0.0091} & \textbf{0.0058} & \textbf{0.0161} & \textbf{0.0079} & \textbf{0.0251} & \textbf{0.0102} & \multicolumn{1}{c}{62.7} & 0.5  & \textbf{0.0151} & \textbf{0.0092} & \textbf{0.0270} & \textbf{0.0130} & \textbf{0.0473} & \textbf{0.0181} & \multicolumn{1}{c}{322.2} & 5.4  \\
    \midrule
    \multicolumn{1}{c|}{Testing} & + TMask-R & \underline{0.0060} & \underline{0.0037} & \underline{0.0109} & \underline{0.0053} & \underline{0.0204} & \underline{0.0076} & -    & 0.6  & 0.0114 & 0.0069 & 0.0207 & 0.0099 & 0.0365 & 0.0139 & -    & 5.6  \\
    \midrule
    \midrule
    \multicolumn{2}{c|}{SASRec (Base)} & 0.0097 & 0.0061 & 0.0160 & 0.0081 & 0.0249 & 0.0104 & -    & 0.5  & 0.0142 & 0.0088 & 0.0253 & 0.0124 & 0.0430 & 0.0168 & -    & 4.5  \\
    \midrule
    \multicolumn{1}{c|}{\multirow{4}[2]{*}{Training}} & + Windows & 0.0088 & 0.0055 & 0.0150 & 0.0075 & 0.0250 & 0.0100 & \multicolumn{1}{c}{87.9} & 0.5  & 0.0147 & 0.0091 & 0.0267 & 0.0131 & 0.0458 & 0.0174 & \multicolumn{1}{c}{360.1} & 4.5  \\
         & + CMR & 0.0110 & 0.0072 & 0.0183 & 0.0094 & 0.0276 & 0.0115 & \multicolumn{1}{c}{74.7} & 0.5  & 0.0155 & 0.0096 & 0.0277 & 0.0135 & 0.0476 & 0.0184 & \multicolumn{1}{c}{394.6} & 4.5  \\
         & + CMRSI & 0.0116 & 0.0080 & 0.0172 & 0.0098 & 0.0268 & 0.0122 & \multicolumn{1}{c}{146.1} & 0.5  & 0.0167 & 0.0101 & 0.0289 & 0.0142 & 0.0490 & 0.0194 & \multicolumn{1}{c}{5611.4} & 4.5  \\
         & + CL4SRec & \textbf{0.0166} & \textbf{0.0112} & \textbf{0.0241} & \textbf{0.0136} & \textbf{0.0359} & \textbf{0.0166} & \multicolumn{1}{c}{105.7} & 0.5  & \textbf{0.0207} & \textbf{0.0130} & \textbf{0.0353} & \textbf{0.0177} & \textbf{0.0584} & \textbf{0.0235} & \multicolumn{1}{c}{958.8} & 4.5  \\
    \midrule
    \multicolumn{1}{c|}{\multirow{2}[2]{*}{Testing}} & + TNoise & \underline{0.0119} & 0.0075 & 0.0195 & 0.0099 & 0.0314 & 0.0129 & -    & 0.6  & \underline{0.0171} & \underline{0.0108} & 0.0293 & \underline{0.0147} & 0.0496 & \underline{0.0198} & -    & 6.1  \\
         & + TMask-R & 0.0117 & \underline{0.0078} & \underline{0.0197} & \underline{0.0101} & \underline{0.0317} & \underline{0.0131} & -    & 0.6  & 0.0170 & 0.0105 & \underline{0.0299} & 0.0146 & \underline{0.0505} & 0.0196 & -    & 6.4  \\
    \bottomrule
    \end{tabular}}}%
  \label{tab:main_result_2}%
\end{table*}%

\subsubsection{Baselines} In addition to the baseline methods already described in Sections \ref{sec:operators} and \ref{sec:existing}, we add a representative training-time augmentation method, \textbf{CLS4Rec} \cite{xie2022contrastive}. It leverages random data augmentation operators (Mask, Crop, and Reorder) with contrastive learning to extract self-supervised signals from the original data, further improving the model performance. In order to verify the generalizability of our method, we validate on more representative models: \textbf{NARM} \cite{li2017neural} incorporates attention into RNN to model the user's sequential pattern. \textbf{NextItNet} \cite{yuan2019simple} combines masked filters with 1D dilated convolutions to model the dependencies in interaction sequences. \textbf{LightSANs} \cite{fan2021lighter} proposes the low-rank decomposed self-attention networks to overcome quadratic complexity and vulnerability to over-parameterization. \textbf{FMLP-Rec} \cite{zhou2022filter} is an all-MLP model with learnable filters for SR tasks. 

\subsubsection{Evaluation Settings} We adopt the leave-one-out strategy to partition sequences into training, validation, and test sets. We rank the prediction over the whole item set rather than negative sampling, otherwise leading to biased discoveries \cite{krichene2020sampled}. The evaluation metrics include Hit Ratio@K (H@K) and Normalized Discounted Cumulative Gain@K (N@K). We report results with K $\in \{5, 10, 20\}$. Generally, \emph{greater} values imply \emph{better} ranking accuracy.

\subsubsection{Implementation Details.} For all baselines, we adopt the implementation provided by the authors. We set the embedding size to 64 and the batch size to 256. To ensure fair comparisons, we carefully set and tune all other hyperparameters of each method as reported and suggested in the original papers. We use the Adam \cite{2014Adam} optimizer with the learning rate 0.001, $\beta_1=0.9$, $\beta_2=0.999$. For TNoise, we tune the $a$ and $b$ in the range of $\{1,0.5,0.05,0.005,0.0005\}$ and $\{2,1,0.1,0.01,0.001\}$, respectively. For TMask, we tune the $\sigma$ in the range of $[0.1,0.9]$ with steps of 0.1. We conduct five runs and report the average results for all methods.

\subsection{Comparison with Testing Augmentation}
The experimental results of based models and adding different test-time augmentation methods are presented in Table \ref{tab:main_result_1}. We also give the time required for model inference when using different methods. We have the following observations and conclusions: 

(1) The performance of the baseline methods are similar to the results in empirical study, i.e., Substitute and Mask are superior. Combinations of multiple operators, such as CMR and CMRSI, still perform worse than using only Mask and Substitute, which suggests that other operators do not work as well when used for TTA. Regarding inference time, Substitute and CMRSI are significantly more time-consuming because of the calculation of item similarity.

(2) TNoise can bring almost no improvement on GRU4Rec, while it can significantly improve SASRec. We speculate that it may be because the RNN-based model is less sensitive to noise than the transformer-based one \cite{lim2021noisy}. It may also be because the attention mechanism models sequences from a global perspective \cite{vaswani2017attention}, whereas RNNs are in left-to-right order, and noise accumulates during the modeling process. TMask-B gives fewer improvements, even slightly lower than the baseline. Setting embedding to 0 may be detrimental to the model's ability to encode and predict user behavior, making a sudden break in the continuous sequence.

(3) TMask-R can achieve the second-best or best performance in all cases. On the Sports and Home datasets, it brings average performance improvement of 72.03\% and 81.28\% to GRU4Rec, 25.27\% and 38.55\% to SASRec, at the cost of just 0.1 minutes of additional inference time. On the Yelp dataset, the average performance gains are 33.74\% and 19.20\%, while adding between 0.2 and 1.9 minutes of extra time. TMask-R introduces appropriate perturbations and higher-relations by removing items while retaining the original sequential patterns and avoiding the interference brought by the mask token, thus achieving optimal performance.

\begin{table}[!t]
  \centering
  \caption{Results on more backbone models. The `Ave. Imp.' represents the average improvement over the base model.}
  \vspace{-1em}
    \renewcommand\arraystretch{0.95}
        \setlength{\tabcolsep}{1mm}{
  \scalebox{0.75}{
    \begin{tabular}{l|cccccccc|c}
    \toprule
     \multicolumn{1}{c|}{\multirow{2}[2]{*}{Method}} & \multicolumn{2}{c}{Beauty} & \multicolumn{2}{c}{Sports} & \multicolumn{2}{c}{Home} & \multicolumn{2}{c|}{Yelp} & \multirow{2}[2]{*}{Ave. Imp.} \\
    \multicolumn{1}{c|}{} & \multicolumn{1}{c}{H@10} & \multicolumn{1}{c}{N@10} & \multicolumn{1}{c}{H@10} & \multicolumn{1}{c}{N@10} & \multicolumn{1}{c}{H@10} & \multicolumn{1}{c}{N@10} & \multicolumn{1}{c}{H@10} & \multicolumn{1}{c|}{N@10} &  \\
    \midrule
    NARM  & 0.0467  & 0.0248  & 0.0299  & 0.0147  & 0.0111  & 0.0055  & 0.0265  & 0.0128  & \multicolumn{1}{c}{-} \\
    + TNoise & 0.0454  & 0.0245  & 0.0301  & 0.0148  & 0.0107  & 0.0053  & 0.0269  & 0.0130  & -0.85\% \\
    + TMask-B & 0.0460  & 0.0247  & 0.0284  & 0.0137  & 0.0110  & 0.0054  & 0.0259  & 0.0127  & -2.44\% \\
    + TMask-R & 0.0479  & 0.0252  & 0.0344  & 0.0178  & 0.0147  & 0.0074  & 0.0279  & 0.0136  & 14.85\% \\
    \midrule
    NextItNet & 0.0364  & 0.0191  & 0.0205  & 0.0102  & 0.0074  & 0.0036  & 0.0187  & 0.0086  & \multicolumn{1}{c}{-} \\
    + TNoise & 0.0369  & 0.0192  & 0.0207  & 0.0103  & 0.0075  & 0.0038  & 0.0192  & 0.0089  & 2.12\% \\
    + TMask-B & 0.0397  & 0.0207  & 0.0267  & 0.0137  & 0.0112  & 0.0053  & 0.0205  & 0.0095  & 25.08\% \\
    + TMask-R & 0.0412  & 0.0208  & 0.0269  & 0.0140  & 0.0113  & 0.0054  & 0.0232  & 0.0111  & 30.80\% \\
    \midrule
    LightSANs & 0.0494  & 0.0274  & 0.0228  & 0.0122  & 0.0118  & 0.0063  & 0.0187  & 0.0089  & \multicolumn{1}{c}{-} \\
    + TNoise & 0.0533  & 0.0291  & 0.0254  & 0.0141  & 0.0135  & 0.0071  & 0.0207  & 0.0096  & 10.84\% \\
    + TMask-B & 0.0515  & 0.0280  & 0.0237  & 0.0125  & 0.0127  & 0.0066  & 0.0196  & 0.0093  & 4.32\% \\
    + TMask-R & 0.0541  & 0.0298  & 0.0289  & 0.0152  & 0.0136  & 0.0072  & 0.0225  & 0.0104  & 17.04\% \\
    \midrule
    FMLP-Rec & 0.0513  & 0.0283  & 0.0247  & 0.0127  & 0.0141  & 0.0075  & 0.0194  & 0.0095  & \multicolumn{1}{c}{-} \\
    + TNoise & 0.0530  & 0.0289  & 0.0256  & 0.0133  & 0.0148  & 0.0079  & 0.0205  & 0.0099  & 4.25\% \\
    + TMask-B & 0.0526  & 0.0287  & 0.0281  & 0.0145  & 0.0145  & 0.0077  & 0.0209  & 0.0103  & 6.69\% \\
    + TMask-R & 0.0576  & 0.0313  & 0.0319  & 0.0168  & 0.0161  & 0.0084  & 0.0234  & 0.0114  & 18.89\% \\
    \bottomrule
    \end{tabular}}}%
  \label{tab:more_models}%
    \vspace{-1em}
\end{table}%

\subsection{Comparison with Training Augmentation}
We compared our methods with training-time augmentation methods. The results are presented in Table \ref{tab:main_result_2}. We also give the retraining time for training-time augmentation methods and inference time for all methods. Regarding recommendation performance, our method performs sub-optimal in almost all cases, outperforming heuristic data enhancement methods such as Sliding Windows, CMR, and CMRSI. CL4SRec utilizes contrastive learning to mine user preferences further and achieve the best performance. However, deploying CLS4Rec requires changing the training step by adding additional learning objectives. In terms of time, deploying our method does not need to retrain a well-trained model. In contrast, CMRSI, which includes the Substitute and Insert operators, requires significant time to complete retraining. TTA methods are still inferior to representative training-time augmentation methods at this stage. In terms of performance alone, TTA methods are still inferior to representative training-time augmentation methods at this stage. However, combining performance, efficiency, and deployment simplicity, our method is competitive and can be used as a free lunch for sequential recommendation models.

\subsection{Generalize to More Backbones}
In order to verify the generalizability of our proposed method, we added our operators to more SR models with different base modules. The results are presented in Table \ref{tab:more_models}. Our approach still effectively improves the model's performance on different SR models, Especially TMask-R. TNoise brings less performance improvement and only exceeds TMask-B in a few cases. Combined with the previous results on GRU4Rec (Table \ref{tab:main_result_1}), we believe that introducing noise for TTA may not be applicable to all models. What can be observed so far is that TNoise can bring more performance gains on self-attention-based SR models. In addition, we observe that TMask-B also brings performance degradation in some cases, and combined with the results in Table \ref{tab:main_result_1}, we believe that directly setting the embedding of certain items to 0 may affect the model's encoding of the sequence. This effect is even more severe than the original interfering information of the mask token. We will continue to explore these in the future. Our TMask-R can bring the most average performance gains of 14.85\%, 30.80\%, 17.04\%, and 18.89\% in all cases, demonstrating its high applicability and generalizability.

\begin{figure}[!t]
  \centering
  \includegraphics[scale=0.55]{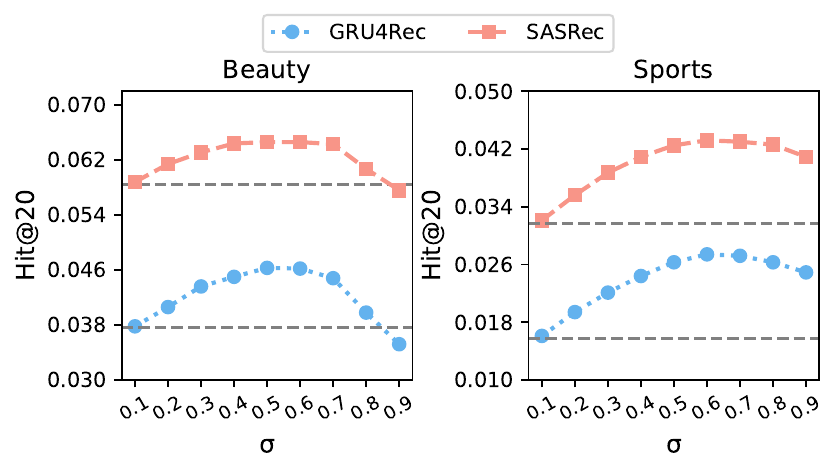}
  \vspace{-1em}
  \caption{Effect of hyperparameter $\sigma$ on performance. The gray line is performance of corresponding base model.}
  \label{fig:sigma}
  \vspace{-1em}
\end{figure}

\begin{table}[!t]
  \centering
  \caption{The performance our method with different hyperparameters. We also give the average similarity between augmented and original data.}
    \vspace{-1em}
    \renewcommand\arraystretch{0.95}
        \setlength{\tabcolsep}{1.5mm}{
  \scalebox{0.85}{
    \begin{tabular}{l|cc|cc|c}
    \toprule
     \multicolumn{1}{c|}{\multirow{2}[1]{*}{Variants}} & \multicolumn{2}{c|}{Beauty} & \multicolumn{2}{c|}{Sports} & \multirow{2}[1]{*}{Ave. Sim.} \\
    \multicolumn{1}{c|}{} & \multicolumn{1}{c}{H@10} & \multicolumn{1}{c|}{N@10} & \multicolumn{1}{c}{H@10} & \multicolumn{1}{c|}{N@10} &  \\
    \midrule
    \multicolumn{1}{l|}{SASRec} & 0.0584  & 0.0310  & 0.0317  & 0.0161  & - \\
    \midrule
    \multicolumn{1}{l|}{+ TNoise(0.005,0.001)} & 0.0586  & 0.0311  & 0.0318  & 0.0162  & 0.9999 \\
    \multicolumn{1}{l|}{+ TNoise(0.05,0.01)} & 0.0585  & 0.0311  & 0.0319  & 0.0162  & 0.9936 \\
    \multicolumn{1}{l|}{+ TNoise(0.5,0.1)} & 0.0627  & 0.0335  & 0.0363  & 0.0186  & 0.8559 \\
    \multicolumn{1}{l|}{+ TNoise(1,0.5)} & 0.0653  & 0.0345  & 0.0422  & 0.0213  & 0.7701 \\
    \multicolumn{1}{l|}{+ TNoise(2,1)} & 0.0618  & 0.0323  & 0.0396  & 0.0204  & 0.7055 \\
    \midrule
    \multicolumn{1}{l|}{+ TMask-R(0.1)} & 0.0588  & 0.0311  & 0.0321  & 0.0161  & 0.9839 \\
    + TMask-R(0.6) & 0.0646  & 0.0344  & 0.0432  & 0.0225  & 0.9308 \\
    + TMask-R(0.9) & 0.0576  & 0.0308  & 0.0409  & 0.0214  & 0.8674 \\
    \bottomrule
    \end{tabular}}}%
  \label{tab:noise}%
\vspace{-1em}
\end{table}%

\subsection{Hyperparameter Investigation}

\noindent \textbf{The $\sigma$ for TMask-R.} Since TMask-R performed best in all cases, we investigated the effect of $\sigma$ on performance. The $\sigma$ determines the ratio of items are removed from the sequence during augmentation. The results are illustrated in Figure \ref{fig:sigma}. The effect of $\sigma$ on performance shows a similar trend across models and datasets. As $\sigma$ increases, the performance gradually improves and decreases after reaching the optimal value. When understood in terms of the strength of the augmentation, smaller values of $\sigma$ produce an augmented sequence that is too similar to the original sequence, while too large a value of $\sigma$ makes the augmented sequence deviate too much from the original data. The right $\sigma$ introduces appropriate perturbations, which in turn improves the final performance of the model.

\vspace{0.3em}

\noindent \textbf{Noise Intensity and Data Similarity.} We investigated the performance of TNoise under different intensities ($a,b$). Following empirical study, we give the average similarity between the augmented and original data throughout the inference. The results are illustrated in Table \ref{tab:noise}. When the noise intensity is too low or too high, the augmented data will either be too similar to the original data or too far off, neither of which will achieve better performance. The model will perform better when the noise intensity is in the middle (i.e., appropriate perturbations). For TMask, we can observe a similar trend. This result complements Table \ref{tab:cos}, i.e., excessive similarity of augmented data will result in little to no performance improvement. In addition, we believe that the strength of appropriate perturbations is different under different methods, i.e., there is a difference in the similarity when optimal performance is achieved.

\vspace{0.3em}

\noindent \textbf{Effect of $m$ on Performance.} For a fair comparison, we set $m = 10$ in all our experiments. Here, we explore the effect of different $m$ on the recommendation performance and present the result in Table \ref{tab:valueofm}. We can observe an overall upward trend in performance with increasing $m$. As each sequence is augmented more times, inference and aggregation of different augmented sequences allow the model to predict the user's behavior better and improve accuracy.

\begin{table}[!t]
  \centering
  \caption{Effect of $m$ on performance with SASRec.}
  \vspace{-1em}
    \renewcommand\arraystretch{0.95}
        \setlength{\tabcolsep}{1mm}{
  \scalebox{0.8}{
    \begin{tabular}{c|c|c|ccccccc}
    \toprule
    \multicolumn{2}{c|}{Value of $m$} & \multicolumn{1}{c|}{Base} & 5     & 7     & 9     & 10    & 11    & 13    & 15  \\
    \midrule
    \multirow{2}[2]{*}{Beauty} & H@10  & 0.0584  & 0.0646  & 0.0640  & 0.0636  & 0.0646  & 0.0652  & 0.0659  & 0.0656  \\
          & N@10  & 0.0310  & 0.0339  & 0.0340  & 0.0339  & 0.0344  & 0.0345  & 0.0349  & 0.0350  \\
    \midrule
    \multirow{2}[2]{*}{Sports} & H@10  & 0.0317  & 0.0418  & 0.0432  & 0.0427  & 0.0432  & 0.0448  & 0.0445  & 0.0451  \\
          & N@10  & 0.0161  & 0.0220  & 0.0221  & 0.0226  & 0.0225  & 0.0233  & 0.0230  & 0.0230  \\
    \bottomrule
    \end{tabular}}}%
  \label{tab:valueofm}%
    \vspace{-1em}
\end{table}%

\section{Related Work}

\subsection{Sequential Recommendation}
Sequential recommendation (SR) is an important task to predict the next item to access based on a sequence of interacted items \cite{zhao2023sequential,zhao2023cross}. Early works leveraged Markov Chains \cite{rendle2010factorizing,he2016fusing} to model user behaviors, where the next item to predict is closely related to the latest few interactions. Later, researchers adopted CNNs \cite{yuan2019simple} and RNNs \cite{2022cmnrec} to capture the relationships among items. For example, GRU4Rec \cite{hidasi2015session} trained a Gated Recurrent Unit (GRU) architecture to model the evolution of user interests. Caser \cite{tang2018personalized} embeds a sequence of recent items in the time and latent spaces, then learns sequential patterns using convolutional filters. Due to the success of the self-attention mechanism \cite{vaswani2017attention}, a series of Transformer-based SR models have been proposed. SASRec \cite{kang2018self} applied the Transformer layer to learn item importance in sequences, which characterize complex item transition correlations. STOSA \cite{fan2022sequential} embeds each item as a stochastic Gaussian distribution and devises a Wasserstein Self-Attention module to characterize item-item relationships. Beyond Transformer-based models, researchers found that filtering algorithms can alleviate the influence of the noise in SR models \cite{zhou2022filter}. They proposed an all-MLP architecture with learnable filters.

\subsection{Data Augmentation for SR}
Data augmentation has been shown to be an effective means of mitigating the data sparsity problem in SR \cite{dang2024repeated,dang2024augmenting}. Early work adopted heuristic methods, such as Sliding Windows \cite{tang2018personalized} and random replacing \cite{wang2021counterfactual}, which split one sequence into many sub-sequences or augment based on counterfactual thinking. Later, some work explored using self-supervised signals to discover more preference knowledge. For example, CL4SRec \cite{xie2022contrastive} constructed contrastive views by three operators, i.e., Crop, Mask, and Reorder. CoSeRec \cite{liu2021contrastive} further improved CL4SRec by presenting two informative operators, Substitute and Insert. In addition to modifying sequences by operators, some work utilizes generative models or specialized modules for augmentation. In this line, DiffASR \cite{liu2023diffusion} adopted the diffusion model for sequence generation. Two guide strategies are designed to control the model to generate the items corresponding to the raw data. ASReP \cite{liu2021augmenting} employed a reversely pre-trained transformer to generate pseudo-prior items for short sequences.

However, deploying all these methods requires retraining, architecture modification, or the introduction of additional parameters. Our operators not only improve the accuracy of the backbone network but also achieve comparable performance to traditional augmentation methods with only a little increase in inference time.

\subsection{Test-Time Augmentation}
Test-time augmentation entails pooling predictions from several transformed versions of a given test input to obtain a final prediction \cite{shanmugam2020and}. It has been widely explored in the field of computer vision, such as image segmentation \cite{wang2019aleatoric} and image classification \cite{kim2020learning,shanmugam2021better}. In addition to the average aggregation of all variants, some researchers also explored the adaptive weights for aggregating \cite{shanmugam2021better,kimura2024test} and automatic search optimal augmentation methods \cite{lyzhov2020greedy}. In graph neural networks, some researchers used TTA to enhance graph representation learning \cite{jin2022empowering} or to enhance the generalization of the model over low-degree nodes \cite{ju2024graphpatcher}. Only one work explored the TTA in recommender systems. TAG-CF \cite{ju2024does} proposed a TTA framework that only conducts message passing once at inference time, effectively utilizing graph knowledge while circumventing most of the computational overheads of message passing. Unlike this effort, we explore and analyze TTA in sequential recommendation from multiple perspectives and propose two new operators.

\section{Conclusion}
In this work, we explore the test-time augmentation for sequential recommendation. We first experimentally reveal the potential of existing data augmentation operators applied to TTA. Further analysis reveals that Mask and Substitute are effective because they retain the original sequential pattern while adding appropriate perturbations. Furthermore, We found that the random selection satisfies the condition of a small number of key interactions and a large number of non-key interactions and achieves a better performance. Meanwhile, Substitutes and Masks suffer from time-consuming item selection and interference by the mask token. Based on the analysis and limitations, we propose two new operators, TNoise and TMask. The former injects uniform noise into the original representation, avoiding the computational overhead. The latter blocks mask tokens from participating in model calculations or directly removes interactions that should have been replaced with mask tokens. Extensive experiments shown that our approach achieves multi-win results in terms of effectiveness, efficiency and generalizability. For future work, we hope to provide theoretical support for our method or propose more stable and efficient methods for TTA.

%%
%% The acknowledgments section is defined using the "acks" environment
%% (and NOT an unnumbered section). This ensures the proper
%% identification of the section in the article metadata, and the
%% consistent spelling of the heading.
\begin{acks}
This work is partially supported by the National Natural Science Foundation of China under Grant No. 62032013, No. 62102074, and the Science and technology projects in Liaoning Province (No. 2023JH3/10200005). We greatly appreciate the anonymous reviewers for their valuable comments and suggestions, which are important for improving and polishing our paper.
\end{acks}

%%
%% The next two lines define the bibliography style to be used, and
%% the bibliography file.
\bibliographystyle{ACM-Reference-Format}
\bibliography{references}

%%
%% If your work has an appendix, this is the place to put it.
% \appendix
% \section{Research Methods}

\end{document}